\documentclass[floatfix,aps,prd,twocolumn,showpacs,10pt]{revtex4}
\usepackage{epsfig}
\usepackage{epsf}
\usepackage{amsmath}
\usepackage{amssymb}

\usepackage{graphicx}

\begin{document}


\preprint{astro-ph/0611851}


\title{Constraining the Variation of $G$  \\ by Cosmic Microwave Background Anisotropies}

\author{K. C. Chan}\email{kcchan@phy.cuhk.edu.hk}
\author{M.-C. Chu}\email{mcchu@phy.cuhk.edu.hk}
\affiliation{	Department of Physics and Institute of Theoretical Physics, the Chinese University of Hong Kong, Hong Kong SAR, PRC.}

\date{\today}


\begin{abstract}
      We use the Cosmic Microwave Background Anisotropies (CMBA) power spectra to constrain the cosmological variation of gravitational constant $G$. It is found that the sensitivity of CMBA to the variation of $G$ is enhanced when $G$ is required to converge to its present value. The variations of $G$  from the CMB decoupling epoch $z \sim 1000$ to the present time are modelled by a step function and a linear function of scale factor $a$ respectively, and the corresponding $95\%$ confidence intervals for $G/ G_0$ are $[0.95, 1.05]$ and $[0.89, 1.13]$, $G_0$ being the present value. The CMBA constraint is unique in the sense that it entails the range of redshift from $z \approx 1000 $ to 0. 

\end{abstract}

\pacs{06.20.Jr, 98.80.Cq, 98.70.Vc}

\maketitle

\section{Introduction}
\label{sect:Intro}
     One of the fundamental questions in physics is whether the fundamental constants are truly constant. Indeed the possibility of cosmological variation of ``constants'' has long been proposed\cite{Dirac}. Among the fundamental constants, the gravitational constant $G$ is the least accurately measured. The value of $G$ is measured in the laboratory and applied to all scales. To check for the constancy of $G$, tests should be done at different spatial and temporal scales. There are many tests at redshifts of order 0. For example, the lunar laser ranging experiment, which monitors the distance between the earth and the moon by laser ranging technique, can be used to put a bound on the variation of $G$\cite{LLR}. If $G$ varies during the history of the earth, the surface temperature and size of the earth would change. However the earth does not preserve a good record of the gravitational conditions. Increase in $G$ causes the Sun to burn at a faster rate, and the depth of the convective zone is affected, which can be observed in the vibrational modes of the Sun, in particular the $p$ mode\cite{Helioseismology}. Since an increase in $G$ also shortens the life spans of stars, the ages of stars in globular clusters can be used to put a constraint on the deviation of $G$ from its present value \cite{GC}. Because the Chandrasekhar mass $M_{Ch} \propto G^{-3/2}m_{N}^{-2}$ sets the mass scale of neutron stars, by observing the masses of neutron stars formed at different redshifts, limits on $G$ in the past 10 Gyr or so have been derived in \cite{Neutron stars}. The highest redshift, of about $10^{10} $, constraint comes from Big Bang Nucleosynthesis (BBN). An increase in the expansion rate during the epoch of BBN causes the freeze-out to occur earlier, and the abundances of neutron and hence ${}^4$He are enriched\cite{BBN}. For details of various experiments and observations, see the review article by Uzan\cite{Uzan constants} (see also \cite{Chiba}). 
         
    On the theory side, there have been grand unification theories and string/M theory motivated models predicting that some of the fundamental ``constants'', such as the fine structure constant $\alpha$ and the Newtonian gravitational constant $G$, may vary over time. In theories with extra dimensions, the effective gravitational constant in 4D spacetime depends on the more fundamental mass scale in the bulk and the size of the extra dimensions\cite{G ED}. If the size of the extra dimensions evolves over time, the effective constants in 4D will vary as well. For example, the DGP model\cite{DGP} has been put forward to explain the recent observation that the universe is in an accelerating phase without invoking the dark energy. It was argued that the acceleration is due to the leakage of gravity into the extra dimensions.  
 
    To select promising ones from the myriads of models in the literature, one may constrain possible variations of the fundamental ``constants'' using observational data. There have been efforts trying to constrain the possible variation of $G$ using cosmological data. In particular, since the Cosmic Microwave Background Anisotropies (CMBA) is sensitive to many cosmological parameters, it could be used to constrain the variation of $G$. CMBA is unique because it offers a long look back time. The physics of CMBA is particularly clean as it involves only well known physics in the linear regime. One approach is to constrain the variation of $G$ in some particular types of models, \textit{e.g.} the CMBA spectra in the Brans-Dicke cosmology are discussed in Ref.~\cite{Chen,Nagata}. However, given the multiplicity of models in the literature, it seems more practical to use a simple and generic parametrization for $G$. In \cite{G_Friedmann, Japanese}, the authors have used the CMBA power spectra to constrain the possible variation of $G$ with a parameter $\lambda$ as
\begin{equation}
 \label{eq:parametrization}
 G=\lambda ^2 G_{0},
\end{equation}
where $G_{0}$ is the present laboratory-measured value. It was assumed that $\lambda$ is a constant over the age of the universe (and only suddenly becomes 1 at the present time). However this assumption is unrealistic since we know that $G$ should converge to its present value to avoid conflicts with other experiments and observations. We shall call the convergence of $G$ to its present value \textit{stabilization}. One can imagine that $G$ may vary in many different manners over the history of the universe, and so it is hopeless to deal with all possibilities one by one. In this article we study two generic stabilization schemes. One of them is instantaneous stabilization. That is we consider an abrupt gravitational transition and model the variation of $G$ by a step function:  
\begin{equation}
\label{eq:parametrization1}
\lambda^2 = \left\{ \begin{array}{ll}
\lambda_{0}^2 & \textrm{if $a< a_s $},  \\
1 & \textrm{if $  a \ge  a_s $ },
\end{array} \right.
\end{equation}
where $a$ is the scale factor and $a_s$ is the scale factor at which stabilization occurs. Another is that $G$ varies smoothly and we parametrize it as a linear function of $a$:
\begin{equation}
\label{eq:parametrization2}
\lambda^2 = \left\{ \begin{array}{ll}
\lambda^2 _{0} & \textrm{if $a< a_* $},  \\
1 - \frac{a_s  -a}{a_s -a_{*}} ( 1- \lambda^2_0 )  & \textrm{if $ a_* \le a \le a_s $ } ,  \\
1 & \textrm{if  $a>a_s$    }, 
\end{array} \right.
\end{equation}
where $a_*$ is the scale factor at the time of photon decoupling. Our main goal is to constrain the range of $\lambda^2_0$ in Eq.~\ref{eq:parametrization1} and Eq.~\ref{eq:parametrization2}. We assume that the underlying mechanism for the variation of $G$ does not affect other physics so that the standard CMBA calculation with the modifications of Eq.~\ref{eq:parametrization}-~\ref{eq:parametrization2} is valid.    
The Friedmann equation is modified as
\begin{equation}
 \label{eq:Friedmann}
  \left( \frac{ \dot{a} }{a^2} \right)^2=\lambda ^ 2(a) H^{2}(a),
\end{equation}
with
\begin{equation}
 H^2(a) = H_{0}^{2} \left(\frac{\Omega_{M}}{a^3} + \frac{\Omega_{\gamma}}{a^4} +  \Omega_{\Lambda} \right),
\end{equation}
where a dot denotes derivative with respect to the conformal time, $H_{0}$ is the present Hubble parameter, $\Omega_{M}$ is the density parameter of the non-relativistic matter, $\Omega_{\gamma}$ is the density parameter of radiation, and $\Omega_{\Lambda}$ is the density parameter of the cosmological constant. Note that we consider a flat universe here. 

     The paper is organized as follows. The effects of variation of $G$ on the CMBA power spectra are investigated in Section~\ref{sect:effect}. In Section~\ref{sect:MCMC}, we study the constraints on the variation of $G$ by the three year WMAP data using the method of Markov Chain Monte Carlo (MCMC) and discuss  the results obtained. Section~\ref{sect:Conclusion} is devoted to the conclusion.

\section{Effect of Variation of $G$ on the CMBA Temperature and Polarization Power Spectra}
\label{sect:effect}

\begin{figure}
	\centering

	\includegraphics[angle=-90,  width=7cm] {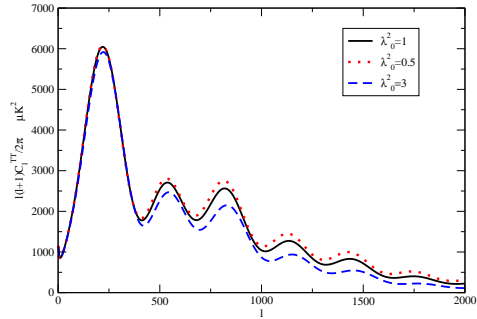}
        \caption{The CMBA temperature angular power spectra for three values of the Newtonian gravitational constant with instantaneous stabilization $z_s=0$. The solid, dotted and dashed curves correspond to $\lambda_0^2=1$, 0.5 and 3 respectively. The damping effect is only noticeable for $l>500$.  }             
	\label{fig:Gspectrum}
\end{figure}

   It has been pointed out in Ref.~\cite{G_Friedmann} that the CMBA angular power spectrum does not change using the simple prescription Eq.~\ref{eq:parametrization} as far as the dynamical equations are concerned. The ratio between the sound horizon and the distance to the last scattering surface (LSS), $\Theta$, will not change if both are blown up by the same factor due to the varied gravitational constant in the flat universe. However, the scaling is not perfect since the recombination physics does introduce another scale. The recombination is dictated mainly by the binding energy of hydrogen atom, which is not affected by a change in the gravitational constant. If the expansion rate is greater during the epoch of recombination, it will be more difficult for the protons and electrons to recombine, and the ionization fraction $x_{e}$ will increase. The probability density that a photon last scatters at a  conformal time $\eta$ is given by the visibility function (for a review of CMBA physics, see for example \cite{Hu Dodelson}),
\begin{equation}
g(\eta)=\dot{\tau} e^{-\tau},
\end{equation}
with 
\begin{equation}
\dot{\tau}=a n_{e} \sigma_{T},
\end{equation}
where $n_{e}$ is the number density of electrons and $\sigma_{T}$ is the Thomson cross-section. The increase in $x_{e}$ broadens $g(\eta)$, resulting in more severe damping of the high $l$ peaks. However the duration that the photons stay in contact with the LSS is also shortened as the expansion rate is greater. Because the two effects partially cancel each other, the damping is not increased much even when $G$ is increased by a large amount. In Fig.~\ref{fig:Gspectrum}, we show the temperature power spectra for $\lambda_0^2 =0.5 $ and 3 compared to the unchanged one ($\lambda_0 ^2 =1$) for the case of of instantaneous stabilization with the stabilization redshift $z_s \equiv \frac{1}{a_s} -1  =0 $. A large change in $G$ is required for noticeable changes in the spectra. The damping effect can be partially compensated by reionization, and so the details of reionization such as the degree of reionization affect the sensitivity of the spectra to $\lambda$.

\begin{figure}
	\centering
	\includegraphics[angle=-90, width=7cm] {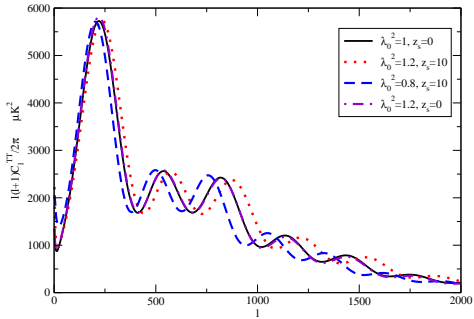}
        \caption{Same as Fig.~\ref{fig:Gspectrum} but with spectra with $z_s=10$ in contrast with those with $z_s=0$. The standard spectrum (solid) nearly coincides with the one with $\lambda_0^2=1.2$ and $z_s=0$ (dot-dashed). When instantaneous stabilization occurs at $z_s=10$, marked shifts to larger (smaller) $l$ scales result for $\lambda_0^2=1.2$ (dotted) (0.8 (dashed)).  }             
	\label{fig:G_varywithZ}
\end{figure}

     The above conclusion that the CMBA angular power spectrum is not sensitive to the value of $G$ is based on the assumption that the gravitational constant remains different from the present laboratory-measured-value  till ``yesterday.'' However $G$ must converge to its present value so that there is no conflict with the low-redshift constraints. When $G$ changes as described in  Eq.~\ref{eq:parametrization1} or Eq.~\ref{eq:parametrization2}, the expansion history of the universe is modified, the fractional change in the sound horizon at the epoch of decoupling is different from that in the conformal distance to the LSS, and the resultant CMBA spectra will be distorted. Fig.~\ref{fig:G_varywithZ} shows the temperature power spectra with instantaneous stabilization at $z_s =0$ and 10. Note that the ``standard'' spectrum ($\lambda_0^2=1$ and $z_s=0$) nearly coincides with the one for $\lambda_0^2 =1.2$ and $z_s=0$. If, however, the stabilization redshift is at $z_s=10$, the spectrum for $\lambda_0^2$=1.2 (0.8) shifts to larger (smaller) $l$ scales. In fact, it is conceptually simpler to compare the one for $\lambda_0^2=1.2$, $z_s=0$ to the one with $\lambda_0^2=1.2$ and $z_s =10$.  The size of the sound horizon is the same for both cases while the distance to the LSS is increased for $z_s=10$; as a result $\Theta$ becomes smaller and the spectrum shifts to high $l$ scales. The same argument applies to the one with $\lambda_0^2=0.8 $ and $z_s=10$, but with the opposite effect. Because of the dramatic gravitational transition, we observe an enhanced late Integrated Sachs Wolfe (ISW) effect in the small $l$ scales.

\begin{figure}
	\centering
	\includegraphics[angle=-90, width=7cm] {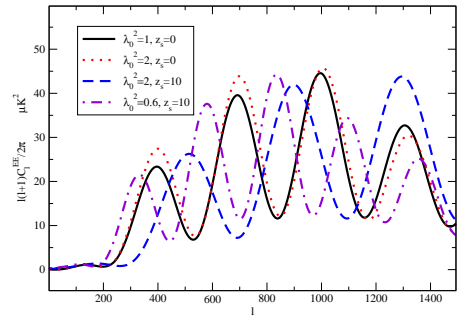}
        \caption{The E-polarization spectra with instantaneous stabilization at $z_s =0$ and 10. The peaks shift to larger (smaller) $l$ scales for $\lambda_0^2=2$ (0.6) when instantaneous stabilization takes place at $z_s=10$, in contrast to the cases without stabilization ($z_s=0 $).}             

	\label{fig:E_G_stab}
\end{figure}

       We observe similar sideway shifts in the E-type polarization and TE cross polarization spectra as well. In Fig.~\ref{fig:E_G_stab}, we show the E-polarization power spectra with instantaneous stabilization at $z_s=0$ and 10. In Ref.~\cite{G_Friedmann}, the authors proposed that the degeneracy between the expansion rate and the scalar spectral index $n_{s}$ can be lifted  by measuring the polarization, because the formation of polarization is proportional to the width of the visibility function. An increase in $G$ causes the power of the polarization spectrum to increase in small $l$ scales and then decrease in large $l$ scales. But the effect of stabilization is much more appreciable.  

\begin{figure}
	\centering
	\includegraphics[angle=-90, width=7cm] {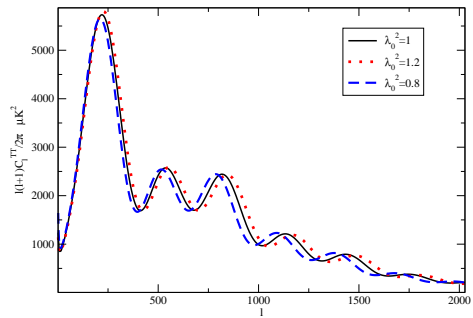}
        \caption{The effects of linear variation of $G$ on the temperature power spectrum. The solid line shows the spectrum when there is no variation of $G$. When $\lambda_0^2=1.2$ (0.8), the peaks shift to larger (smaller) $l$ scales. $a_s$ is set to be 1.}  

	\label{fig:GLinear}
\end{figure}

        When $G$ varies linearly with $a$, its effects on the CMBA power spectra are still dominated by the sideway shifts discussed in the instantaneous stabilization scenario, as can be seen in Fig.~\ref{fig:GLinear}. The effects are qualitatively similar for the two stabilization schemes that we use.

    Since our prescription is simple, we can easily estimate the amount of shift in the CMBA spectrum due to stabilization. This calculation is similar to that of the CMB shift parameter\cite{Melchiorri}. From Eq.~\ref{eq:Friedmann}, we have the conformal distance to the LSS
\begin{eqnarray}
d_{*} \equiv \eta_{0} - \eta_{*}  = \int _{ a_{*} } ^{1} \frac {d a}{a^2 \lambda (a) H(a)},  
\end{eqnarray}
where $\eta_0$ is the present conformal time, and $\eta_{*}$ is the conformal time at CMB decoupling. One immediately sees that if $\lambda_0$ is larger, the conformal distance will be smaller. This is reasonable because it takes less time for the universe to expand to its present size. The peak positions in the CMBA temperature angular power spectrum in a flat universe can be characterized by \cite{Hu Dodelson}
\begin{equation}
\label{eq:ldr_relation}    
    l_{n}\approx n \pi \frac{d_{*}}{r_{s}},
\end{equation}
where $r_s$ is the sound horizon: 
\begin{equation}
r_s =  {\int_{0}^{\eta_{*}} c_{s} d \eta}. 
\end{equation}
The sound speed $c_s$ is given by 
\begin{equation}
\label{eq:cs}
    c_{s}^2=\frac{1}{3(1+R)},
\end{equation}
with
\begin{equation}
    R=\frac{3}{4} \frac{\rho_{B}}{\rho_{\gamma}}, 
\end{equation}
where $\rho_{B}$ and $\rho_{\gamma}$ are mass densities of baryons and radiations respectively. It should be pointed out that the sound horizon also depends on $\lambda$ through $\eta$. Hence if $\lambda_0>1$ at the epoch of CMB decoupling, the size of the sound horizon will also be smaller. In fact, the reduction rates for $d_{*}$ and $r_{s}$ are the same if there is no stabilization; the effects of $\lambda$ are exactly cancelled, a manifestation that the peaks do not shift if there is no stabilization. However, when there is stabilization, two different scales will be introduced and the peak positions will shift.
      
\begin{figure}
	\centering
	\includegraphics[width=9cm] {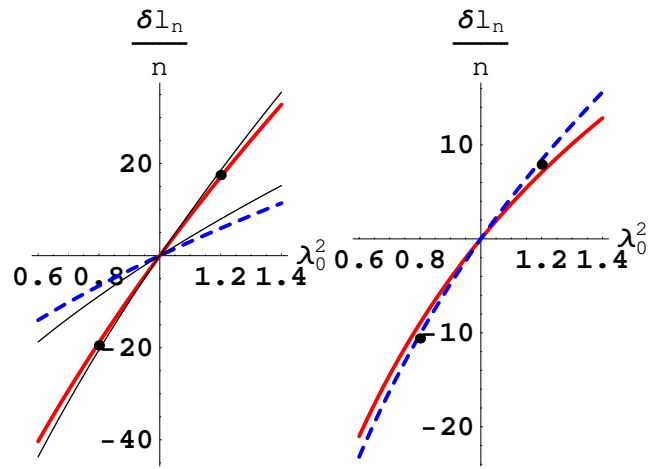}
        \caption{  The left (right) panel shows $\delta l_n /n  =  \pi \delta(d_{*} / r_s )$ vs. $\lambda_0^2$ for instantaneous (linear) stabilization. For the plot on the left, the thick (red) solid curve corresponds to $a_s$=0.1 and the dashed curve to $a_s$=0.5. The dots correspond to $\delta l_n / n$ averaged over the first three peaks of the spectra in Fig.~\ref{fig:G_varywithZ} ($\lambda_0^2 =1.2, z_s=10$ and $\lambda_0^2=0.8, z_s=10$). The two thin solid (black) curves are obtained using the simplified formula Eq.~\ref{eq:simplifiedEq} for $a_s=1$ and 0.5 respectively. On the right, the solid and dashed curves correspond to linear stabilization  with $a_s=1$ and 0.5 respectively. Similarly we denote by the dots the average $\delta l_n /n $ in Fig.~\ref{fig:GLinear} for $a_s=1$, and $\lambda_0^2$=1.2 and 0.8 respectively. 
 }             
	\label{fig:SemiAnalytic_shift}
\end{figure}

     We now illustrate with the case of instantaneous stabilization. The calculations can be simplified by noting that if there is no stabilization, the peaks do not shift even if $\lambda_0^2 \neq 1$. The sound horizon is the same irrespective of stabilization if it takes place after decoupling. Hence we only need to compare the conformal distance between the case with stabilization and the one without it:
\begin{eqnarray}
   \label{eq:deltad} 
   \delta d_{*} & = &  d_{*NS} - d_{*S}      \\
   & = &  \int _{a_{s} }^ {1}  \left( \frac{1}{\lambda_{0}} -1 \right) \frac {d a}{a^{2} H(a)}, 
\end{eqnarray}
where the subscript $_{NS}$ denotes no stabilization and $_{S}$ denotes with stabilization. From Eq.~\ref{eq:ldr_relation}, the shifts in the peak positions due to the change in $G$ are given by 
\begin{eqnarray}
\label{eq:shiftl}
   \delta l_{n} 
   &=& \frac{ n \pi} {r_{s}} \delta d_{*}.   
\end{eqnarray} 
Plugging in the standard $\Lambda$CDM model parameters, we have
\begin{equation}
\label{eq:simplifiedEq}
\delta l_n  \sim  508 n (\lambda_0 -1)  (0.558- \sqrt{ 0.311 a_s }  ).
\end{equation}
One may proceed similarly for linear variation of $G$, but it is too cumbersome to write down the results explicitly. We display $\delta  l_n   / n =\pi \delta(d_{*} / r_s )$ against $\lambda_0^2$ in Fig.~\ref{fig:SemiAnalytic_shift}. We see that the shift in $l$ from the simple arguments here agrees with the full numerical calculations in Fig.~\ref{fig:G_varywithZ} and Fig.~\ref{fig:GLinear}. Furthermore, the curve by the simplified formula in Eq.~\ref{eq:simplifiedEq} tracks closely the one from complete calculations of $r_s$ and $d_*$.   

\section{Constraining $G$ by MCMC and Discussions}
\label{sect:MCMC}
      Because the CMBA angular power spectra are sensitive to various parameters, to be consistent, other relevant parameters should also be varied when fitting with data. A popular means is to make use of the Markov Chain Monte Carlo (MCMC) method. With the temperature and polarization spectra computed by the Boltzmann code CMBFAST\cite{CMBFAST}, we employ the public MCMC engine CosmoMC\cite{CosmoMC} to explore the parameter space. Since the Hubble parameter $H_{0}$ is measured to rather good precision by the HST key project, we take $H_{0}=72$\cite{Freedman}. The constraints we get are not sensitive to this restriction. Thus the free parameters that we vary are: $\omega_{B}=\Omega_{B}h^2$, $\omega_{CDM}=\Omega_{CDM}h^2$, $z_{re}$, the reionization redshift, $n_{s}$, the index of the primordial perturbation spectrum, $A_{s}$, the normalization amplitude, $\lambda_0^2$, and the stabilization redshift $z_s$. We use the three year WMAP data\cite{WMAP3yr} to constrain these parameters.

\begin{figure}
	\centering
        \includegraphics[width=7cm] {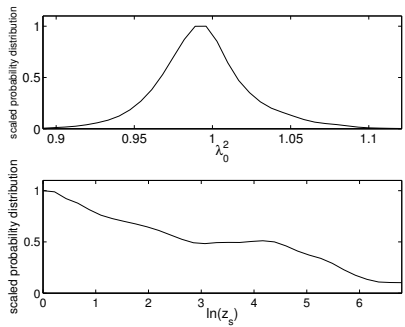}
        \caption{The marginal distributions of $\lambda_0^2$ and $\ln z_s$ obtained using the three-year WMAP data set in the instantaneous stabilization scenario. Here and thereafter, the maximum of the distribution is normalized arbitrarily to 1. }             
	\label{fig:Lambda2_Zs_dis}
\end{figure}

\begin{figure}
	\centering
	\includegraphics [width=7cm] {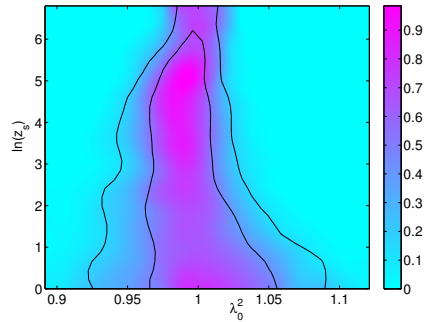}
        \caption{The contour plot of the joint distribution of $\lambda_0^2$ and $\ln z_s $ constrained using the three-year WMAP data set in the instantaneous stabilization scenario. The inner and outer solid lines are the 68\% and 95\% confidence level contours respectively.}             
	\label{fig:Lambda2_Zs_cont}
\end{figure}

     First of all, we do not consider stabilization; that is, we assume that $z_{s}=0$. Imposing the prior that $\lambda_0^2 <2.2$, we get the constraint on $\lambda_0^2$ to be [0.91, 2.20] at 95\% confidence level. The bounds seem to be sensitive to the prior on $G$. The weak constraint on $G$ is expected given the small change in the CMBA power spectra even for relatively large variations in $G$ as discussed in Section~\ref{sect:effect}.  Now we implement the instantaneous stabilization parametrized by Eq.~\ref{eq:parametrization1}. Since there are already tight constraints on the variation of $G$ at redshifts of order 0, we impose the prior of $\ln z_s > 0$. On the other hand, we want to constrain the variation of $G$ after CMB decoupling, and thus we impose the prior that $\ln z_s < 6.8 $.    The marginal distributions of $\lambda_0^2$ and $ \ln z_s$ are shown in  Fig.~\ref{fig:Lambda2_Zs_dis}. The $2\sigma$ confidence intervals of $\lambda_0^2$ and $ \ln z_s$ are [0.95, 1.05] and [0, 5.57] respectively. With stabilization, the confidence interval of $\lambda_0^2 $ shrinks substantially. In Fig.~\ref{fig:Lambda2_Zs_cont}, we show the contour plot of the joint marginal distribution of $\lambda_0^2$ and $ \ln z_s$. The triangular shape of the distribution is due to the fact that the constraint on $\lambda_0^2$ is tighter if $\ln z_s$ is larger. 

\begin{figure}
	\centering
	\includegraphics [width=7cm] {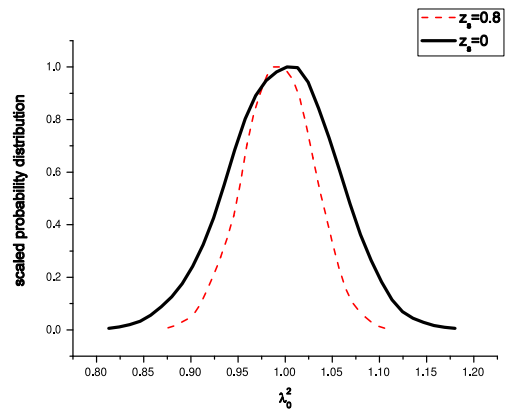}
        \caption{The marginal distributions of $\lambda_0^2$ for the linear stabilization scenario in Eq.~\ref{eq:parametrization2} with  $z_s=0$ and  $z_s=0.8$ respectively, constraint using the three-year WMAP data. }      
        \label{fig:GLinear_dis}
\end{figure}

We now turn to the linear stabilization given in Eq.~\ref{eq:parametrization2}.
For smooth variation, we set $z_s=0$. The marginal distribution of $\lambda_0^2$ for $z_s=0$ is shown in Fig.~\ref{fig:GLinear_dis}, and the resultant 95\% confidence interval of $\lambda^2_0$ is [0.89,1.13].

 Translating the above constraint on the linear parametrization of $G$ to the common form $ \dot{G}/G $, we get $ \dot{G}/G = (-9.6 \sim  +8.1 ) \times 10^{-12}$ yr$^{-1}$. This is complementary to the constraints from neutron star mass and BBN, which constrain the variation of $G$ in the regimes of redshifts $0 \sim  4$ and $10^{10}$  respectively\cite{Chiba}. The results are summarized in  Table~\ref{tab:constraint_table}. 
\begin{table}[h!]
\caption{The constraints on the variation of $G$ at various redshifts. The CMBA constraint fills the gap in the ``redshift ladder'' in between neutron star mass and BBN.} 
\label{tab:constraint_table}
\begin{tabular}{|l|l|l|}
\hline
{   }    & redshift &  $\dot{G}/G $ {   }   yr$^{-1} $   \\
\hline
Lunar laser ranging \cite{LLR}   & 0 & $(1 \pm 8 ) \times 10^{-12} $  \\
\hline
Neutron star mass \cite{Neutron stars}   & $0 \sim  4 $ & $(-0.6 \pm2.0) \times 10^{-12} $   \\
\hline
CMBA (WMAP)  & $ 0 \sim 1000 $  & $ (-9.6 \sim 8.1 ) \times 10^{-12} $  \\
\hline
BBN \cite{BBN}   & $ 10^{10} $   &  $ (-2.7 \sim 2.1) \times 10^{-11} $    \\

\hline

\end{tabular}

\end{table}
We see that the CMBA power spectra can extend the constraint on the variation of $G$ to a large range of redshifts that other experiments and observations cannot reach. Improving the low-redshift bounds on $G$ helps to tighten the bounds at high redshifts because we can set $z_s$ to a large value. This is supported in Fig.~\ref{fig:Lambda2_Zs_cont} for instantaneous stabilization and also for linear stabilization in Fig.~\ref{fig:GLinear_dis}, where we also plot the marginal distribution of $\lambda_0^2$ with $z_s=0.8$. The allowed range of $\lambda_0^2 $ is smaller than that of $z_s=0$.  

      Although we only consider two types of parametrizations, they in fact encompass a large class of models in which $G$ varies monotonically after photon decoupling. If $G$ varies sharply near some redshift $z_s$, this can be approximated by the parametrization Eq.~\ref{eq:parametrization1}, and the contour distribution in Fig.~\ref{fig:Lambda2_Zs_cont} can be applied. On the other hand if $G$ varies in a smooth manner, the linear parametrization Eq.~\ref{eq:parametrization2} can be a good approximation.

     We note that other cosmological parameters that we also vary agree well with the standard $\Lambda$CDM cosmological parameters given in \cite{WMAP_para}, and so we do not bother to write them down. Without stabilization, all cosmological parameters are still within $2 \sigma$ from those in the WMAP paper. The agreement gets better when we take stabilization into account. In these cases, all are within $1\sigma$. 
     
      From the MCMC runs, we can analyze the degeneracy between $\lambda_0^2$ and other cosmological parameters. We see that $\lambda_0^2$ has some degeneracy with $\omega_B$, $\omega_{CDM}$ and $n_s$. The degeneracy with $\omega_B$ and $\omega_{CDM}$ can be understood by the fact that $\lambda_0^2$ and $\omega_i$ ($i=B$ or $CDM$) appear in the Friedmann equation as the product $\lambda_0^2 \omega_i$. The change in relative amplitudes on different scales due to $\lambda_0^2$ can be compensated by changing the relative amplitudes in the primordial perturbations characterized by $n_s$ \cite{G_Friedmann}. As mentioned earlier, for instantaneous stabilization, $\lambda_0^2$ is quite strongly degenerate with $\ln z_s$. It is well-known that the effect of curvature on the spectrum is to shift it sideways, somewhat similar to the effect of stabilization of $G$, which is the most important contribution to our constraints. Thus we expect that our bounds on $\lambda_0^2$ will be weakened by the inclusion of curvature in the fitting.

     Since the upcoming Planck satellite mission is going to probe the temperature power spectrum to as high as $l \sim  2500$ and the E-polarization spectrum to $l \sim 1500$, we expect that there will be tremendous improvement in the constraint on $\lambda_0^2$. We can forecast the improvement that Planck will bring us quantitatively using the Fisher matrix, which has been widely used to predict the expected uncertainties in future experiments (see \textit{e.g.} \cite{G_Friedmann, Eisenstein}). Under the assumption of Gaussian perturbations and Gaussian noise, the Fisher matrix takes the form
\begin{equation}
F_{i j} = \sum_{l} \sum_{X,Y} \frac{\partial C _{Xl} }{\partial p_i}  (\mathrm{Cov}_{lXY})^{-1} \frac{\partial C_{Yl} }{\partial p_j},
\end{equation}
where $p_i$ is the $i$th free parameter and $C_{X l}$ is the $l$th multipole of the observed spectrum of type $X$, which can be the temperature, temperature-polarization and E-polarization spectra. The experimental precision is encoded in the covariant matrix $\mathrm{Cov}_{lXY}$. We find that with the temperature power spectrum alone the current constraint is improved by a factor of 8; when the polarization spectrum is included, the bounds will be tightened by a factor of 11 relatively to our current bounds. The CMBA constraints on $\dot{G}/G$ will be potentially one of the best constraints. Furthermore, it is possible to strengthen the constraint by including the matter power spectrum as well since enhanced ISW effects are induced by the gravitational stabilization.

\section{Conclusion}
\label{sect:Conclusion}

     Previously, CMBA was used to constrain the variation of $G$ without considering stabilization. Not only are the resultant limits weak, but also this assumption does not respect many tight local constraints. In this work we consider two simple and generic parametrizations of $G$, the instantaneous stabilization and linear stabilization. Stabilization causes appreciable sideway shifts in the CMBA power spectra, and hence the sensitivity of CMBA to the variation of $G$ is enhanced. We use the three year WMAP data to constrain the model parameter(s) and other cosmological parameters. For the case of instantaneous stabilization, we simultaneously constrain $\lambda_0^2$ and $\ln z_s$ to [0.89, 1.13] and [0, 5.57] to  $2\sigma$ intervals. For the linear stabilization scenario, $z_s$ is set to 0 and we get the 95\% confidence interval [0.89, 1.13], which is equivalent to $\dot{G}/G = (-9.6 \sim 8.1 ) \times 10^{-12}$ yr$^{-1} $. Although we only concentrate on two types of parametrization, our results can be applied to a large class of models in which $G$ varies monotonically after CMB decoupling because the variation of $G$ in many of these models can be approximated by either a step function or a linear function.  The constraint derived from CMBA extends the bounds on $G$ up to the redshift of about 1000, and so it is complementary to other experiments and observations. In particular, it fills the ``redshift gap'' between BBN and the neutron star mass constraints. With the forthcoming Planck data, the constraints will be improved by a factor of 10 or so, and the constraints on $\dot{G}/G$ from CMBA may be one of the best ones.

\begin{acknowledgments}
We are grateful for K. Umezu for useful communications. We also thank the ITSC of the Chinese University of Hong Kong for using its clusters for calculations. This work is partially supported by a grant from the Research Grant Council  of the Hong Kong Special Administrative Region, China (Project No. 400803).    
\end{acknowledgments}


\end{document}